\documentclass[aps,pra,nofootinbib,,eqsecnum,preprint,
showpacs,amsmath,amssymb,floatfix,showpacs]{revtex4}
\usepackage{graphicx}

\begin{document}

\title{Casimir friction: Relative motion more generally}         

\author{Johan S. H{\o}ye  and Iver Brevik}        

\address {Department of Physics, Norwegian University of Science and Technology, 7491 Trondheim, Norway,}
\address{Department of Energy and Process Engineering, Norwegian University of Science and Technology, 7491 Trondheim, Norway}

\begin{abstract}

This paper extends our recent study on Casimir friction forces for dielectric plates moving parallel to each other [J. S. H{\o}ye and I. Brevik, Eur. Phys. J. D {\bf 68}, 61 (2014)], to the case where the plates are no longer restricted to rectilinear motion. Part of the mathematical formalism thereby becomes more cumbersome, but reduces in the end to the form that we could expect to be the natural one in advance. As an example,  we calculate the Casimir torque on a planar disc rotating with constant angular velocity around its vertical symmetry axis next to another plate.

\end{abstract}
\maketitle

\bigskip
\section{Introduction}
\label{sec1}

In this work we will continue our study of the Casimir friction force between two dielectric plates (half-spaces) that move longitudinally with respect to each other with a small separation $d$. See Fig.~1.

\begin{figure}[b]
\includegraphics[width=2.5in]{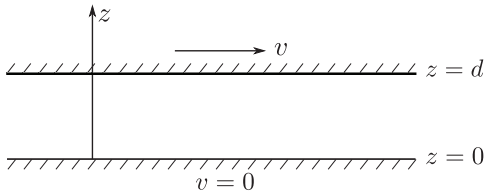}
\caption{Standard configuration: Upper plate moving with velocity $\bf v$; lower plate at rest. Gap width is $d$.}
\end{figure}

This is a topic that has attracted considerable interest in the recent past. In a recent paper of ours \cite{hoye14} a problem of this sort was analyzed: we calculated the Casimir friction for constant velocity, at zero temperature as well as at finite temperature, requiring vanishing initial and final velocities in order to obtain a closed loop motion meaning a return to the starting position. The friction force was found  via the dissipated energy by which the net contribution from the slow velocity part could be neglected and thus did not require further specification. Moreover, we assumed the simple situation with a constant velocity $v$ between the times $-\tau$ and $\tau$. In addition we assumed very low initial and final velocities in the opposite directions in order to be able to return to the starting position.

In the present work we want to extent our results to the situation where the finite velocity contributing to dissipation is not restricted to be constant, but may be slowly varying, and not necessarily restricted to rectilinear motion. By that, circular motion with constant speed can be considered too.  A nice feature of the latter kind of motion, besides constant speed, is its return to the initial position as required by the energy dissipation method.

On physical grounds it is reason to expect that with slowly varying velocity the total dissipated energy will be the sum of contributions from the various velocities. This is provided the constant velocity case considered in Ref.~\cite{hoye14} lead to a correct result. However, this extension of the problem is non-trivial. The reason is that somehow contributions from different velocities have to be separated from each other while in the reference the contributions from the very slow initial and final velocities could be neglected anyway. As will be seen in Sec.~\ref{sec2} to facilitate this separation of contributions, we find it necessary to split the integrand of Eq.~(\ref{2}) below in two terms. These terms are subdivided in different time intervals as given by Eq.~(\ref{4A}).

For the more general motion considered here, it actually turns out that essentially all the derivations and results of Ref.~\cite{hoye14} remain unchanged. The exception is the integral containing the specified motion, which becomes more cumbersome and requires a detailed and more accurate treatment to handle nonzero contributions for different velocities.

Some papers dealing with  Casimir friction - most of them quite new - are listed in Refs.~\cite{teodorovich78,pendry97,pendry98,pendry10,volokitin99,volokitin03,volokitin07,volokitin08,volokitin11,dedkov08,dedkov10,
dedkov11,dedkov12,philbin09,barton10A,barton10B,barton11,barton11B,silveirinha13A,maghrebi13,intravaia14,pieplow13,kyasov14,henkel14}. Here one will find studies also for the case where one single particle is traveling close to a dielectric surface. In principle, the theory of systems of this kind can be  obtained from the full theory of interacting dielectric planes, in the limit of large dilution for one of the planes.

It should be noted that the formalism we use to obtain the friction force has been developed by us in previous works starting with a pair of polarizable particles as a basis and computing the response via the Kubo formalism \cite{kubo59, hoye92,hoye10,hoye11a,hoye11b,hoye12a,hoye12b,hoye13}. The methods used by us are quite different from the approaches used by others referred to above.

The bases for the use of the Kubo formula are developments and results obtained in the statistical mechanics of polar and polarizable fluids. Via the Feynman path integral this formalism was extended to polarizable particles whose oscillations were quantized \cite{feynman53,hoye81,chandler82}. This formalism was further extended by the authors to evaluate Casimir forces \cite{brevik88} and to evaluate Casimir friction \cite{hoye92, hoye93}. Then it turned out that time dependent interactions like the radiating dipole interaction could be included too in the statistical mechanical treatment where imaginary time is the forth dimension. An advantage of this formalism is that the electromagnetic field can be disregarded (or eliminated). Instead it is replaced by dipolar interactions between pairs of polarizable particles. Another advantage is thus  the possibility to consider media on microscopic level where particles are separated by a minimum distance due to molecular hard cores. In this way it is possible to evaluate (approximately) the finite Casimir energy in bulk of a simple fluid model \cite{waage13}. With the statistical mechanical approach the Casimir forces may be given an alternative physical interpretation; they are induced molecular attractions due to fluctuating dipole moments.

 In the next section dealing with rectilinear motion where the velocity may vary, the key point will be how to handle properly the  $\hat Q$ integral (Eq.~(\ref{2}) below) and the product with its complex conjugate. This is needed in order to obtain  the total dissipated energy. Under the present general circumstances there will be contributions to the dissipation from various velocities that can not be neglected. In Ref.~\cite{hoye14} the simplifying situation with only one velocity was regarded as the contribution from the slow initial and return motions can be neglected anyway.  After carrying out this more careful analysis, we consider two-dimensional motion in the horizontal plane in Section III. Finally, we consider in Section IV as an example, a rotating planar disc above a resting plate and evaluate the Casimir torque.

 As before, we find the friction force to be proportional to $v^3$  at $T=0$, assuming $v$ small, while it is proportional to $v$ at finite $T$, assuming $v$ small.

\section{Rectilinear motion}
\label{sec2}

 As in Ref.~\cite{hoye14} we consider a two-plate setup in which the lower plate (2) is at rest, while the upper plate (1) executes motion in a closed loop meaning that it finally slides back to its initial position. The friction force is evaluated via the dissipation of energy, the latter point being advantageous since one avoids the problem of separating a reversible part of the inter-particle force from the total force.

 Let us outline some essentials of the theory given in Ref.~\cite{hoye14} keeping out details of evaluation that can be found there. For simplicity the numeral I will be used below to designate the equations of Ref.~\cite{hoye14}.  So consider a quantum mechanical harmonic two-oscillator system whose Hamiltonian $H$ is perturbed by a time-dependent term written in general form as $-AF(t)$. Her $A$ is a time independent operator and $F(t)$ is a classical function that depends upon time. For simplicity consider for the moment a pair of one-dimensional oscillators for which we can write
\begin{equation}
 -AF(t)=-\psi({\bf r}(t))s_1 s_2 \label{1a}
 \end{equation}
where $\bf r$ is the separation between the pair of oscillators, $\psi(\bf r)$ is the coupling strength, and $s_1,s_2$ are the vibrational coordinates of the oscillators. The instantaneous force between the oscillators is ${\bf B}=-\nabla \psi({\bf r})s_1s_2$.
  Its thermal average is, according to the Kubo formula given by Eq.~(I3)
  \begin{equation}
  \langle {\bf B}(t)\rangle =\int_{-\infty}^t\phi_{BA}(t-t')F(t')dt'. \label{A}
  \end{equation}
 With
 \begin{equation}
 A=s_1 s_2 \quad \mbox{and} \quad F(t)=-\psi({\bf r}(t)) \label{C}
 \end{equation}
the response function is
 \begin{equation}
 \phi_{BA}(t)=\frac{1}{i\hbar}{\rm Tr}\left\{\rho [A,{\bf B}(t)]\right\}. \label{B}
 \end{equation}
 Here $\rho$ is the density matrix and
  ${\bf B}(t)=e^{itH/\hbar}{\bf B}e^{-itH/\hbar}$ is the Heisenberg operator.
 Further we can write the response function as
 \begin{equation}
 \phi_{BA}(t)=\nabla \psi \,\phi(t), \label{D}
 \end{equation}
 where
 \begin{equation}
 \phi(t)=\frac{1}{i\hbar}{\rm Tr}\left\{ \rho [s_1s_2,s_1(t)s_2(t)]\right\}. \label{E}
 \end{equation}
The $\phi(t)$ depends upon the temperature and the polarizabilities, $\alpha_1$ and $\alpha_2$, and the eigenfrequencies, $\omega_1$ and $\omega_2$, of the two oscillators as given by Eqs.~(I18)-(I22).
 \begin{equation}
\phi(t)=C_-\sin(\omega_-t)+C_+\sin(\omega_+t),
 \label{D1}
 \end{equation}
 \begin{equation}
C_\pm=\frac{H}{\hbar}\sinh(\frac12\beta\hbar\omega_\pm), \quad H=\frac{\hbar^2\omega_1\omega_2\alpha_1\alpha_2}{4\sinh(\frac12\beta\hbar\omega_1)\sinh(\frac12\beta\hbar\omega_2)}
 \label{D2}
 \end{equation}
with $\omega_\pm=|\omega_1\pm\omega_2|$ ($\phi(t)=0$ for $t<0$) and $\beta=1/(k_B T)$ where $T$ is temperature and $k_B$ is Boltzmann's constant.
The relative position between the two oscillators can be written as
\begin{equation}
 {\bf r}= {\bf r}_0+{\bf v}q(t).\label{E1}
 \end{equation}
 With $\dot{F}(t)=-({\bf v\nabla }\psi)\dot{q}(t)$, we can write the dissipated energy for fixed ${\bf r}_0$ as
 \begin{equation}
 \Delta E({\bf r}_0)=-\int_{-\infty}^\infty {\bf v}\dot{q}(t)\langle {\bf B}\rangle dt=-\int_{-\infty}^\infty \int_{-\infty}^t \dot{F}(t)\phi(t-t')F(t')dt'dt. \label{F}
 \end{equation}

For two half-planes with surfaces located at $z=0$ and $z=d$ one can for low densities integrate to obtain
 the total energy dissipation per unit surface as
\begin{equation}
\Delta E=\rho_1 \rho_2\int\limits_{z_1>d,z_2<0} \Delta E({\bf r}_0)\,dx_1dy_1dz_1dz_2,
\label{G}
\end{equation}
where $\rho_1,\rho_2$ are the uniform number densities. We write this as
\begin{equation}
\Delta E=\rho_1\rho_2\int\limits_{t>t'} L(t,t')\phi(t-t')dtdt', \label{H}
\end{equation}
and find after some calculation by use of Fourier transform methods that $L(t,t')$ takes the form of Eq.~(I12)
\begin{eqnarray}
\nonumber
L(t,t')&=&-\int \dot F(t)F(t')\,dx_1dy_1dz_1dz_2\\
&=&-\frac{1}{(2\pi)^2}\int\limits_{z_1>d,z_2<0}\hat\psi(z_0,{\bf k}_\perp)\hat\psi(z_0,-{\bf k}_\perp)A(t,t')\,d{\bf k}_\perp dz_1dz_2,
\label{I}
\end{eqnarray}
where
\begin{equation}
A(t,t')=-i{\bf k}_\perp {\bf v}\dot q(t) e^{-i{\bf k}_\perp {\bf v}(q(t)-q(t'))}
\end{equation}
with  $d{\bf k}_\perp=dk_xdk_y$ and $z_0=z_1-z_2$.

The $A(t,t')$ is to be integrated together with the $\phi(t-t')$ of Eq.~(\ref{D1}). With use of the condition of return $q(\infty)=q(-\infty)$ (=0) one finds that it can be rewritten as Eq.~(I17)
\begin{equation}
A(t,t')=\frac{1}{2}\sum_{n=\pm1}\dot Q(t,n\omega_v)Q(t',-n\omega_v),
\label{K}
\end{equation}
where
\begin{equation}
Q(t,\omega_v)=e^{-i\omega_v q(t)}-1,\quad \mbox{with}\quad \omega_v={\bf k}_\perp {\bf v}=k_xvq(t).
\label{M}
\end{equation}

Then the dissipated energy becomes expression (I23) which is
\begin{equation}
\Delta E=\frac{\rho_1\rho_2}{(2\pi)^3)}\int\limits_{z_1>d, z_2<0}\hat\psi(z_0, {\bf k}_\perp)\hat\psi(z_0, -{\bf k}_\perp)J(\omega_v)\,d{\bf k}_\perp dz_1 dz_2
\label{N}
\end{equation}
\begin{equation}
J(\omega_v)=\int\limits_{t>t'} A(t,t')\phi(t-t')\,dt dt'=C_- I(\omega_-)+C_+ I(\omega_+)
\label{O}
\end{equation}
The $C_\pm$ are the coefficients given by Eq.~(\ref{D2}). For higher densities straightforward summation (or integration) of particle pairs is no longer valid due to dipolar interactions within each half-plane. This, however, is taken into account by replacing the polarizability $\alpha=\alpha(\omega)$ by the corresponding dielectric constant $\epsilon$. The replacement is
 $2\pi\rho\alpha\rightarrow(\varepsilon-1)/(\varepsilon+1)$ as given by Eq.~(I50). This extension to arbitrary densities we showed in Sec.~4 of Ref.~\cite{hoye13}.

By some calculation one finds Eq.~(I26)
\begin{equation}
I(\omega)=\frac{\omega}{4}\sum_{n=\pm1}\hat Q(-\omega,n\omega_v)\hat Q(\omega,-n\omega_v).
\label{P}
\end{equation}

So far, the formalism works out similarly as in the previous case of Ref.~\cite{hoye14}. The new element in our analysis is to calculate
 the integral (I28) of Ref.~\cite{hoye14}. This integral is
\begin{equation}
\hat Q(\omega,-\omega_v)=\int\limits_{-\infty}^\infty(e^{i\omega_v q(t)}-1)e^{-i\omega t}\,dt.
\label{2}
\end{equation}
 With $q(t)$ consisting only of a few linear parts in $t$, the  integral (\ref{2}) is easily evaluated. But to obtain the appropriate form of the result was less trivial as the product of $\hat Q$ and its complex conjugate in Eq.~(I26) or Eq.~(\ref{P}) should produce the $\delta$-functions of Eq.~(I29). But the corresponding $\delta$-functions for the slow initial and return motions were not considered as the dissipation should vanish anyway for these parts. For the present situation with varying velocity all finite velocities will contribute and thus can not be neglected. So to obtain the desired result in this more general situation the difference between the two terms of the integral have to be taken in a proper way.

If the velocity $v_x=v\dot q(t)$ varies slowly the $q(t)$ can be considered piecewise linear in $t$ such that explicit integrations can be performed. However, the additional problem is that expression (\ref{2}) should be multiplied with its complex conjugate as mentioned above by which cross-terms will appear. The problem is to get rid of these cross-terms. As will be seen below this is possible by separating the integrand in two parts that are subdivided differently in intervals.

 Then consider a time interval from $t_1$ to $t_2$ of length $2\tau=t_2-t_1$. These times are chosen as limits for part of the first term of the integral of Eq.~(\ref{2}). The corresponding interval for the second term of the integral is chosen from $t_1'$ to $t_2'$ such that
\begin{eqnarray}
\nonumber
\omega t_1'&=&\omega t_1-\omega_v q(t_1)\\
\omega t_2'&=&\omega t_2-\omega_v q(t_2)
\label{4A}
\end{eqnarray}
With this subdivision of the two terms of the integrand the full integral will be covered properly by such intervals when the motion that starts at time $t_s$ ends at the same position at time $t_e$, i.e.~the condition $q(t_s)=q(t_e)=0$ is fulfilled.

Relation (\ref{4A}) can now be expanded around the middle of the time intervals. So to linear order with $t_1=t_0-\tau$, $t_2=t_0+\tau$, $t_1'=t_0'-\tau'$, and $t_2'=t_0'-\tau'$ condition (\ref{4A}) becomes
\begin{eqnarray}
\nonumber
\omega (t_0'-\tau')=\omega(t_0-\tau)-\omega_v(q(t_0)-\dot q(t_0)\tau)\\
\omega (t_0'+\tau')=\omega(t_0+\tau)-\omega_v(q(t_0)+\dot q(t_0)\tau)
\label{5}
\end{eqnarray}
from which follows
\begin{equation}
\omega\tau'=(\omega-\omega_v\dot q(t_0))\tau \quad \rm{and} \quad \omega t_0'=\omega t_0-\omega_v q(t_0).
\label{6}
\end{equation}

For the chosen interval one now gets the integrals (with $x=t'-t_0'$ and then $x=t-t_0$
\begin{eqnarray}
\nonumber
S_1&=&\int\limits_{t_1'}^{t_2'}e^{-i\omega t'}\,dt'=e^{i\omega t_0'}\int\limits_{-\tau'}^{\tau'} e^{-i\omega x}\,dx=2e^{i\omega t_0'}\frac{\sin{(\omega\tau')}}{\omega}\\
&=&2e^{i\omega t_0'}\frac{\sin{((\omega-\omega_v\dot q)\tau)}}{\omega},
\label{7}
\end{eqnarray}
\begin{eqnarray}
\nonumber
S_2&=&\int\limits_{t_1}^{t_2}e^{-i(\omega t-\omega_vq(t))}\,dt=e^{i\omega t_0}\int\limits_{-\tau}^\tau e^{-i(\omega-\omega_v\dot q) x}\,dx\\
&=&2e^{i\omega t_0'}\frac{\sin{((\omega-\omega_v\dot q)\tau)}}{\omega-\omega_v\dot q}.
\label{8}
\end{eqnarray}
Here the relations of Eq.~(\ref{6}) are utilized and $\dot q(t_0)=\dot q$ is used as simplification. From this the contribution to integral (\ref{2}) becomes
\begin{equation}
\Delta \hat Q(\omega, -\omega_v)=S_2-S_1=2e^{-i\omega t_0'}\frac{\omega_v \dot q \sin{(\omega-\omega_v\dot q)\tau}}{\omega(\omega-\omega_v\dot q)}.
\label{9}
\end{equation}
According to Eq.~(\ref{P}) this should be multiplied with its complex conjugate to obtain the following contribution
\begin{equation}
\Delta I(\omega)=\sum_{n=\pm 1}\frac {(\omega_v\dot q)^2}{\omega}\left(\frac{\sin{(\omega-\omega_v\dot q)\tau}}{\omega-\omega_v\dot q}\right)^2.
\label{10}
\end{equation}
For large $\tau$ ($\rightarrow\infty$) $\delta$-functions are obtained with amplitude determined by the integral $\int_{-\infty}^\infty(\sin x/x)^2\,dx=\pi$. Thus for large $\tau$
\begin{equation}
\Delta I(\omega)=\pi\tau\frac{(\omega_v\dot q)^2}{\omega}[\delta(\omega-\omega_v\dot q)+\delta(\omega+\omega_v\dot q)].
\label{11}
\end{equation}
With $\dot q=1$ this is Eq.~(I29).

Likewise there will be similar contributions from the other time intervals of the motion. When adding these contributions to Eq.~(\ref{9}) they will form cross-terms when multiplied together. However, products of terms for different time interval with midpoints $t_{10}'$ and $t_{20}'$ will have a phase factor $e^{\pm i\omega(t_{20}'-t_{10}')}$. This phase factor will vary rapidly as function of $\omega$ since $|t_{20}'-t_{10}'|$ can be chosen large when $q(t)$ is slowly varying. So from this argument we find that cross-terms with such phase factors should vanish by the further integrations of $\omega$ and $\omega_v$. With the lack of cross-terms contributions like the ones of Eq.~(\ref{11}) will add such that Eq.~(I29) is modified into ($dt_0=2\tau$)
\begin{equation}
I(\omega)=\int\Delta I(\omega)\frac{dt_0}{2\tau}=\int\frac{\pi(\omega_v\dot q)^2}{2\omega}[\delta(\omega-\omega_v\dot q)+\delta(\omega+\omega_v\dot q)]\,dt_0.
\label{12}
\end{equation}

So altogether with varying velocity the various velocities give independent and additive contributions to the dissipation. With two eigenfrequencies one has only $\omega=\omega_1\pm\omega_2$. In the general situation one has bands of eigenfrequencies and integrations of $I(\omega)$ are performed as in Ref.~\cite{hoye14} starting with Eq.~(I33) to obtain the resulting dissipation.

\section{Motion in the plane}
\label{sec3}

The results in the previous section are for rectilinear motion.  However, it can be extended to more general motion in a straightforward way. Without relative rotation the motion is then such that the term ${\bf v}q(t)$ of Eq.~(\ref{E1}) is replaced by
\begin{equation}
x=x(t)=vq_x(t), \quad y=y(t)=vq_y(t).
\label{13}
\end{equation}
But integral (\ref{2}) can be kept where now
\begin{equation}
\omega_v q(t)\rightarrow k_x x+k_y y.
\label{14}
\end{equation}
As before the velocity is expected to vary slowly to be considered approximately constant within a long time interval $2\tau$. Expanding around its midpoint $t=t_0$ we have with $u=t-t_0$ (with $q_x=q_x(t_0)$ etc.)
\begin{equation}
x=vq_x+v\dot q_x u,\quad y=vq_y+v\dot q_y u, \quad \omega_v q\rightarrow k_x v q_x+k_yvq_y+\omega_v\dot q u
\label{15}
\end{equation}
where now $\omega_v={\bf kv}$ with
\begin{eqnarray}
\nonumber
v_x&=&v\cos\varphi_v=v\frac{\dot q_x}{\dot q}, \quad v_y=v\sin\varphi_v=v\frac{\dot q_y}{\dot q}, \quad \dot q^2=\dot q_x^2+ \dot q_y^2,\\
\omega_v&=&{\bf kv}=kv\cos(\varphi_k-\varphi_v), \quad k_x=k\cos\varphi_k, \quad k_y=k\sin\varphi_k.
\label{16}
\end{eqnarray}
Integral (\ref{2}) can now be performed as before, and for its two terms condition (\ref{4A}) will be modified to

\begin{eqnarray}
\nonumber
\omega t_1'&=&\omega t_1-(k_xv q_x(t_1)+k_yv q_y(t_1))\\
\omega t_2'&=&\omega t_2-(k_xv q_x(t_2)+k_yv q_y(t_2)).
\label{4}
\end{eqnarray}
Likewise expansion (\ref{5}) can be used and conditions (\ref{6}) are still valid with the minor replacement $\omega_v q(t_0)\rightarrow k_xv q_x(t_0)+k_yv q_y(t_0)$. with this, all remaining results (\ref{7}) - (\ref{12}) are still valid.

However, there might be a remaining problem as the velocity changes direction by which the angle $\varphi_v$ of Eq.~(\ref{16}) will vary slowly with time. But this will not influence remaining integration with respect to {\bf k} when following the derivations in Ref.~\cite{hoye14} since only the relative angle between {\bf k} and {\bf v} will occur anyway.

Altogether, we have found that the result for energy dissipation and friction obtained in Ref.~\cite{hoye14} is valid for more general motion. Plates that move relative to each other in a closed circle with only one constant speed $\dot q=$const. will be such a situation.

\section{Rotating planar disc}
\label{sec4}

The results obtained in Sec.~\ref{sec3} will be valid for more general motion where the plates also can rotate with respect to each other. Such a situation will be pure rotation around a center at constant angular velocity. See Fig.~1, where now the upper plate (radius $R$) rotates with  angular velocity $\Omega$ around the vertical axis $z$. The lower plate is at rest, and is of infinite extent, as before.

The argument is that a rotating plate can be subdivided in small areas whose linear dimension is large compared to the separation from the plate at rest. Each area can thus be regarded as a macroscopic plate that moves around. This latter small area will also perform a rotation. But since its linear size is much smaller than that of the whole plate, this rotation contributes to negligible differences between velocities within each small area by which they can be considered equal. Thus for each of them the results of Sec.~\ref{sec3} are valid. This is at least obvious for low dielectric constant in which case the resulting friction force is the sum of contributions for each separate particle.

For a rotating plate it is of interest to have the torque acting due to friction. For two metal plates of the same material at temperature $T=0$ the friction force per unit area in Ref.~\cite{hoye14} was by its Eq.~(I56) found to be
\begin{equation}
F_P=-C_P v^3,\quad C_P=\frac{15\pi^2}{64d^6}\rho^2 D^2\hbar^3,\quad D=\frac{\hbar\nu}{\rho(\pi\hbar\omega_p)^2}
\label{20}
\end{equation}
with dielectric function $\varepsilon=1+\omega_p^2/(\xi(\xi+\nu))$ where $\xi=i\omega$. (Here only small frequencies $\omega\ll\omega_v$ in the corresponding frequency distribution were needed.) The $\rho$ is the particle density of free electrons, $\omega_p$ is the corresponding plasme frequency, and $d$ is the separation between the plates.

Likewise at finite temperature $T$, the corresponding friction force was by Eq.~(59) in the reference found to be
\begin{equation}
F=-Cv,\quad C=\frac{\pi^4}{4\beta^2d^4}\rho^2D^2\hbar
\label{21}
\end{equation}
with $\beta=1/(k_B T)$ where $k_B$ is Boltzmann's constant. Here $d/(\beta\hbar v)\gg1$ is assumed which holds unless $T$ is very small or $v$ is very large.

To obtain results (\ref{20}) and (\ref{21}) the frequency distribution for the dielectric function given below (\ref{20}) for both metal half-planes is needed. It is found via the imaginary part of this function and is given by Eq.~(I51) (for small $m$)
\begin{equation}
m^2\alpha_I(m^2)=Dm, \quad m=\hbar\omega.
\label{21a}
\end{equation}
The $\alpha_I(m^2)$ with $m=m_1$ and $m=m_2$ respectively replaces the product of polarizabilities $\alpha_1$ and $\alpha_2$ in the $C\pm$ given by Eq.~(\ref{D2}). With this replacement the $J(\omega_v)$ and thus the  $\delta$-functions of $I(\omega_\pm)$ are integrated with volume element $d(m_1^2)d(m_2^2)$. Then for $T=0$ only the $C_+$ term contributes while for finite $T$ only the $C_-$ term contributes as $\omega_v$ inside the $\delta$-functions then can be neglected. This results in Eqs.~(I64) and (I52) respectively for $J(\omega_v)$. Then its $\omega=k_x v$ dependence is averaged over directions. Further the electrostatic dipolar interaction has to be i inserted in Eq.~(\ref{N}). This is obtained from the Coulomb interaction $\psi=\psi(r)$ by which the corresponding dipolar interaction $\psi_{ij}$ ($i,j=x,y,z$) is given by Eq.~(I36) as
\begin{equation}
\psi_{ij}=-\frac{\partial^2}{\partial x\partial y}\psi,\quad \psi=\frac1r.
\label{21b}
\end{equation}
The Fourier transforms in the $xy$-plane are
\begin{equation}
\hat\psi_{ij}(z_0,{\bf k}_\perp)=-k_i k_j\hat\psi(z_0,{\bf k}_\perp), \quad \hat\psi(z_0,{\bf k}_\perp)=\frac{2\pi e^{-q|z_0|}}{q}
\label{21c}
\end{equation}
where here $q=k_\perp$, $k_\perp^2=k_x^2+k_y^2$, $ik_z=\mp q$ (for $z_0\gtrless0$). The $\hat\psi_{ij}$ (with $\sum_{ij}$) substitutes the $\hat \psi$ in Eq.~(\ref{N}) to obtain Eq.~(I41)
\begin{equation}
\hat\psi(z_0,{\bf k}_\perp)\hat\psi(z_0,-{\bf k}_\perp)\rightarrow\hat G(z_0,q)=(2q^2)^2\left(\frac{2\pi e^{-q|z_0|}}{q}\right)^2.
\label{21d}
\end{equation}
Finally the integrations of Eq.~(\ref{N}) are performed to obtain results (\ref{20}) and (\ref{21}) above with $\rho_1=\rho_2=\rho$.

For the torque on a rotating plate to be finite it should have a finite radius $R$. With this the torque due to friction for metal plate rotating with angular speed $\Omega$ at $T=0$ will be (with $v=\Omega r$)
\begin{equation}
\tau_P=\int\limits_0^RrF_P2\pi r\,dr=-2\pi C_P\int\limits_0^R\Omega^3r^5\,dr=-\frac{\pi}{3}C_P R^6\Omega^3.
\label{22}
\end{equation}
Likewise for finite temperature the torque will be
\begin{equation}
\tau=\int\limits_0^RrF2\pi r\,dr=-2\pi C\int\limits_0^R\Omega r^3\,dr=-\frac{\pi}{2}C R^4\Omega.
\label{23}
\end{equation}
As noted above Eq.~(\ref{P}) these results for metal plates, with dielectric function given below Eq.~(\ref{20}), are not restricted to a pairwise approximation for pairs of particles, but is valid for arbitrary densities.

Here it can be noted that the $T>0$ result (\ref{21}) (apart from a small factor $\approx 1.2$) agrees with a result obtained earlier by Volokitin and Persson \cite{volokitin07} as shown in Ref.~\cite{hoye13}. Further in Ref.~\cite{hoye14} we showed that the $T=0$ result (\ref{20}) agrees with the one obtained by Barton (except for the factor $\zeta(5)=1.037$) \cite{barton11}. Except for a numerical factor 2 (or 12) it is in accordance with an earlier result by Pendry \cite{volokitin07,pendry97}. In this respect, however, our results, like those mentioned, are not in agreement with the recent ones of Silveirinha  \cite{silveirinha13A}. There, for instance, the quantum friction force is expected to have exponential growth, but is mentioned to be consistent with the semi-classical result of Pendry \cite{pendry97,pendry10} in the weak interaction limit. Also a velocity threshold above which quantum friction can take place was found in Ref.~\cite{silveirinha13A}. We can see no such threshold as the friction is present for all velocities. This reference also draws conclusions about relativistic velocities where Cherenkov radiation will appear. We, however, can not draw such conclusions about Cherenkov radiation as we use electrostatic dipole interaction (\ref{21b}) and thus assume non-relativistic velocities.

In a recent work a freely rotating disc or cylinder was considered \cite{maghrebi14}. This, however, is a situation quite different form the one considered in this work with a disc or plate rotating close to a another parallel plate. Also we limit ourselves to the electrostatic field (near field) while friction on a freely rotating cylinder or disc requires energy loss by radiation. Thus for various reasons our results can not be compared to those of this recent reference.

\section{Summary}
\label{sec5}

We have extended our previous results for Casimir friction to the situation where the velocity may vary both in magnitude and direction. As might be expected we find that the various velocities give independent contributions to the dissipated energy. In Ref.~\cite{hoye14} our results were compared with those of others both for temperatures $T=0$ and $T>0$, and agreement with results of Refs.~\cite{pendry97}, \cite{volokitin07}, and \cite{barton11} were (mainly) obtained.


\end{document}